# Ultrahigh-*Q* AlGaAs-on-insulator microresonators for integrated nonlinear photonics

Weiqiang Xie,[1,†,*] Lin Chang,[1,†] Haowen Shu,[1,2,†] Justin C. Norman,[3] Jon D. Peters,[1] Xingjun Wang,[2] and John E. Bowers[1,3]

[1]*Department of Electrical and Computer Engineering, University of California, Santa Barbara, CA 93106, USA*
[2]*State Key Laboratory of Advanced Optical Communications System and Networks, Department of Electronics, School of Electronics Engineering and Computer Science, Peking University, Beijing, 100871, China*
[3]*Institute for Energy Efficiency, University of California, Santa Barbara, CA 93106, USA*
*Corresponding author: weiqiangxie@ucsb.edu*

**Abstract**

**Aluminum gallium arsenide (AlGaAs) and related III-V semiconductors have excellent optoelectronic properties. They also possess strong material nonlinearity as well as high refractive indices. In view of these properties, AlGaAs is a promising candidate for integrated photonics, including both linear and nonlinear devices, passive and active devices, and associated applications. For integrated photonics low propagation loss is essential, particularly in nonlinear applications. However, achieving low-loss and high-confinement AlGaAs photonic integrated circuits poses a challenge. Here we show an effective reduction of surface-roughness-induced scattering loss in fully etched high-confinement AlGaAs-on-insulator nanowaveguides, by using a heterogeneous wafer-bonding approach and optimizing fabrication techniques. We demonstrate ultrahigh-quality AlGaAs microring resonators and realize quality factors up to $3.52\times10^6$ and finesses as high as $1.4\times10^4$. We also show ultra-efficient frequency comb generations in those resonators and achieve record-low threshold powers on the order of ~20 μW and ~120 μW for the resonators with 1 THz and 90 GHz free-spectral ranges, respectively. Our result paves the way for the implementation of AlGaAs as a novel integrated material platform specifically for nonlinear photonics, and opens a new window for chip-based efficiency-demanding practical applications.**

## 1. INTRODUCTION

Aluminum gallium arsenide (AlGaAs) has rich optical and electrical properties [1] which has been widely used in optoelectronic devices such as laser diodes and photodetectors operating near its bandgap. On the other hand, when used with photon energies far below its bandgap, e.g., in the communication bands, AlGaAs is particularly of great interest in nonlinear fields due to several advantages. First, AlGaAs offers both the strong second- and third-order nonlinearity [2-4]. Second, AlGaAs has a relatively large bandgap, which can be further tailored by varying Al content in the alloy $Al_xGa_{1-x}As$ [5], meaning that a wide transparency window from near- to mid-infrared is achievable. By engineering the bandgap, nonlinear losses like two-photon absorption can be significantly minimized. Finally, AlGaAs has a large refractive index [6] (e.g., $n \approx 3.27$ at 1550 nm for $x = 0.2$), potentially enabling strong confinement for photons. While efficient nonlinear processes in AlGaAs waveguides on native GaAs substrates have been reported in the last decade [7-10], nonlinear studies on an AlGaAs-on-insulator platform have gained increasing interests over the past few years [11-18], including for efficient second harmonic generation, super-continuum generation, and Kerr frequency comb generation. In such a platform, an AlGaAs nanowaveguide for light guiding and interaction is integrated on a low-index silicon dioxide layer using a wafer-bonding approach, and thus those advantages in AlGaAs can be fully exploited, ultimately allowing high-confinement, ultrahigh-efficiency, compact nonlinear photonic integrated circuits (PICs). In integrated nonlinear photonics, low propagation loss is essential for lowering the operating power and improving efficiency in nonlinear processes, for instance, a high-quality ($Q$) factor in a microresonator can substantially reduce the power threshold in frequency comb generation. In fact, in the last few years, we have witnessed the great success of improving nonlinear efficiency by reducing waveguide loss in various nonlinear material platforms such as silica [19], silicon nitride [20-23], and lithium niobate [24-26]. On those nonlinear platforms, significant advances have been made to increase $Q$-factors in microresonators and consequently have enabled high-efficiency on-chip nonlinear applications such as the sub-milliwatt threshold for frequency comb generation and chip-based soliton microcombs operated at a few milliwatts. Likewise, for the implementation of AlGaAs in integrated nonlinear photonics, the key lies in the realization of low loss in nanowaveguides. However, it is likely to be more challenging to achieve low-loss PICs in the AlGaAs-on-insulator platform, primarily due to high-index contrast and large scattering loss compared to dielectric material platforms. While providing strong mode confinement, a high-index contrast between the waveguide core (with an index of $n_{core}$) and cladding (with an index of $n_{clad}$) can significantly increase surface-roughness-induced scattering loss, which is proportional to $(n_{core}^2 - n_{clad}^2)$ [27-29]. For instance, this value in AlGaAs is ~4.5 times higher than that in silicon nitride when the cladding is the oxide. As a result, the propagation loss is very sensitive to waveguide surface roughness, which is more distinct in fully etched nanowaveguides. Therefore, an effective path to lower propagation loss is to reduce the surface roughness.

In this work, we use a heterogeneous wafer-bonding technique [13] to integrate an epitaxially grown $Al_xGa_{1-x}As$ (here $x=0.2$ with a bandgap of ~1.69 eV) crystalline film onto a Si substrate. By systematically optimizing fabrication processes including etching, photoresist (PR) reflow, and AlGaAs growth and bonding, we are able to dramatically reduce the surface roughness on both waveguide sidewalls and bottom/top surfaces. In waveguide fabrication, we investigate two types of epitaxial AlGaAs, which are grown by metal-organic chemical vapor deposition (MOCVD) and molecular-beam epitaxy (MBE), respectively. It is found that, while high-quality waveguides can be made from both epitaxial approaches, the MBE AlGaAs performs better than the film grown by MOCVD in terms of material quality and waveguide loss. With MBE AlGaAs, we demonstrate ultrahigh-$Q$ AlGaAs microrings with a fully etched 400 nm thickness and realize $Q$ factors up to $3.52 \times 10^6$, corresponding to a propagation loss as low as 0.17 dB/cm. In these microrings, cavity finesses as high as $1.4 \times 10^4$ are also attained. To the author's knowledge, the $Q$ factor and finesse reported here are the highest values in III-V semiconductor PICs, which are also on par with the state-of-the-art results in silicon PICs based on a silicon-on-insulator technology. Moreover, as a demonstration of the capability of high nonlinear efficiency, we show ultra-efficient Kerr frequency comb generation in the microring resonators and achieve record-low threshold powers of ~23 µW and ~120 µW for the rings with a 1 THz and a 90 GHz free-spectral range (FSR), respectively. Combining high nonlinearity and low loss in the AlGaAs PICs, our result paves the way for the implementation of AlGaAs as a novel integrated material platform specifically for nonlinear photonics, and opens a new window for chip-based efficiency-demanding practical applications.

## 2. DESIGN AND FABRICATION

In our demonstration, a 400 nm thick AlGaAs film is employed for waveguide fabrication, and both epitaxial films, grown by commercially available MOCVD and in-house MBE, are investigated. Microring resonators are used to extract $Q$ factors and waveguide loss, which have different radii and widths from 600 nm to 1000 nm in design. When the width is below ~800 nm, the waveguides exhibit anomalous group-velocity dispersion at around 1550 nm [17] and are used for demonstration of a Kerr frequency comb generation. The coupling between the bus waveguide and microring is carefully designed to allow efficient coupling only for the fundamental transverse-electric (TE) mode. The width of the waveguide at the facet is designed to be 200 nm to reduce the coupling loss to ~3-4 dB per facet with a typical lensed fiber. Figure 1(a) shows the simulated fundamental TE mode of a 700 nm wide AlGaAs waveguide on SiO$_2$, with an effective mode area ($A_{\text{eff}}$) of only 0.23 µm$^2$, due to the strong mode confinement. To illustrate the superior nonlinear characteristics in such an AlGaAs waveguide, we can calculate the nonlinear waveguide parameter $\gamma$ defined as $\gamma = 2\pi n_2/\lambda A_{\text{eff}}$ with $n_2$ being nonlinear refractive index [30]. Compared to other waveguide systems, e.g., silicon nitride with a typical $A_{\text{eff}}$ of ~1.0 µm$^2$ and $n_2$ two orders of magnitude lower than AlGaAs [11, 31], the nonlinear coefficient in AlGaAs waveguides is more than 400 times higher than that in silicon nitride waveguides.

The fabrication of the AlGaAs waveguide includes two primary steps: wafer bonding and waveguide patterning. The objective of the wafer bonding process is to transfer the epitaxial AlGaAs film grown on a GaAs substrate onto a silicon substrate with a 3 µm thick thermal SiO$_2$ box layer on top. In our experiment, a 3-inch AlGaAs epitaxial wafer and 4-inch silicon wafer are used. This step consists of bonding of the AlGaAs epitaxial wafer and GaAs substrate and etch-stop layer removal (see Supplement 1). Note that the as-grown surface of AlGaAs will be the bottom surface of the waveguide, while the interface between AlGaAs and the etch-stop layer will be the waveguide top surface. Thus, the AlGaAs surface quality by growth and the choice of the etch-stop layer and its removal process are critical for the final waveguide surfaces and loss. The roughness of both surfaces was monitored during fabrication by using atomic force microscope (AFM). The patterning of waveguide involves lithography and dry etching processes. Here we opted for deep-ultraviolet (DUV) lithography with an illumination wavelength of 248 nm, which is better suited for high-level integration and mass fabrication. The PR patterns during lithography and the dry etching of AlGaAs are critical for the final waveguide sidewalls, in which a smooth and vertical surface is highly desired for both low loss and good geometry control. In our fabrication, an oxide hard mask was used for etching AlGaAs, and in particular, a two-step-etch process was developed for the hard mask (see Supplement 1), which can avoid any organic residues and unwanted pattern transfers in the final waveguides. The dry etching for AlGaAs was also optimized (see Supplement 1). Figure 1(b) shows the scanning electron microscope (SEM) image of the cross section of the fabricated AlGaAs waveguide, in which a vertical sidewall is clearly seen. Figure 1(c) shows a waveguide-coupled AlGaAs microring, with the coupling region enlarged in Figure 1(d), indicating a clean surface and a good pattern definition in our process. Lastly, it should be mentioned that during the whole fabrication, all the surfaces of the AlGaAs waveguide were passivated with a thin (5-10nm) $Al_2O_3$ layer by atomic layer deposition (see Supplement 1) for assistance in bonding and possible reduction of the absorption caused by surface states [32].

## 3. *Q* MEASUREMENT RESULTS

The next fabrication step is to clad the waveguide with a ~1.5 μm oxide. The *Q* factors were characterized by measuring the transmission spectra of waveguide-coupled microrings. The TE polarized light from a tunable laser was coupled into the bus waveguide through a lensed fiber and the output was collected by another lensed fiber. The transmitted power was detected with a high-speed photodetector and the spectra were recorded using an oscilloscope. The spectral resolution of our measurement is 0.01 pm. The input power was attenuated low enough to suppress thermal effects. For the measurement of frequency comb generation, the same tunable laser was used as a pump, and comb spectra were recorded using an optical spectrum analyzer.

In this section, we focus on surface roughness reduction in AlGaAs waveguides via fabrication optimization and show its effect on *Q*-factor improvement in AlGaAs microresonators. We first use MOCVD epitaxial AlGaAs in our experiment and then employ MBE-grown AlGaAs for waveguide fabrication using optimized processes. Finally, we present a summary of the results and discussions.

### A. Sidewall Roughness Reduction

As discussed above, the surface roughness in high-index contrast AlGaAs waveguides, specifically the root-mean-square (RMS) roughness ($\sigma$), can cause significant scattering loss, which is proportional to $\sigma^2$ [27]. Among all waveguide surfaces, the sidewall roughness, due to lithography and etching, is always an important contribution, and our first attempt is to reduce such roughness. An effective way to attain this is reflowing PR which can result in a smooth PR pattern edge (see Supplement 1), ultimately avoiding roughness transfer into waveguides during etching. The reflow of PR should be carried out carefully, since it may cause rounding of PR patterns and finally sloped etched sidewalls. In our case, thanks to high etching selectivity (>10) of AlGaAs over $SiO_2$, the hard mask can be very thin to eliminate such a rounding effect.

In the investigation of the sidewall roughness effect on waveguide loss, we adopted commercial MOCVD AlGaAs with an etch-stop layer of 500 nm InGaP between AlGaAs layer and GaAs substrate. With this epitaxial wafer, the waveguides have a roughness of $\sigma \approx 0.45$ nm for bottom surface and $\sigma \approx 0.75$ nm for top surface (see Supplement 1). In Fig. 2, we show the results of etched AlGaAs waveguides without and with the PR reflow process. From the high-resolution top-down SEM images (Fig. 2(a) and 2(c)), it is clearly seen that the line-edge roughness of waveguide is significantly reduced. For a quantitative comparison, we estimated the roughness $\sigma$ from the line-edge roughness profile extracted from the SEM images and reveal that the roughness $\sigma$ has deceased from ~2.2 nm without reflow to ~0.9 nm with reflow, implying a remarkable reduction of sidewall roughness. The effect of reflow is even manifested in the sidewall SEM images (Fig. 2(b) and 2(d)), which show a clear trench-like roughness transferred from PR patterns without reflow, while a smooth sidewall with the reflow process. To examine the effect of the reflow on waveguide loss, we fabricated AlGaAs microrings and tested $Q$ factors. For a ring with a radius of 30 μm and a width of ~850 nm, the intrinsic $Q$-factor was measured to be ~$0.2\times10^6$ without the reflow and up to ~$1.0\times10^6$ with the reflow, implying ~5 times reduction of scattering loss. This also suggests that the scattering loss due to sidewall roughness in the waveguide without reflow process is dominant. In the following experiments, the reflow process was applied for all waveguide fabrication.

**B. Bottom and Top Surface Roughness Reduction**

Since the roughness of the bottom/top surfaces in the waveguide is still high and even comparable to that of the sidewalls, our second attempt for the reduction of scattering loss is to reduce the roughness of the bottom/top surfaces. Here we replaced the etch-stop layer of InGaP with $Al_{0.8}Ga_{0.2}As$ in the MOCVD growth, which is commonly used as an etch stop layer for selective etching of GaAs [33]. Using an optimized etch-stop removal process (see Supplement 1), we achieved a roughness of $\sigma$ less than 0.25 nm for both the bottom and top surfaces, as shown in Fig. 3(a) and 3(b). Here the bottom is the as-grown surface of AlGaAs by MOCVD and the top is the surface after the removal of the etch-stop layer.

We fabricated the microrings and measured the intrinsic $Q_0$ of up to $2.2\times10^6$ in the ring of similar dimension already demonstration, as shown in Fig. 3(c). Cleary, the $Q$ factor has increased from $1.0\times10^6$ to $2.2\times10^6$ after the improvement of the surfaces. The highest $Q$ was measured in a microring with a larger radius of 143 μm and a wider width of 1000 nm, with $Q_0 = 2.72\times10^6$, as shown in Fig. 3(d). Given the fact that the sidewall roughness now dominates over the bottom/top roughness, the major contribution of waveguide loss is caused due to scattering by sidewalls. Nevertheless, higher $Q$ factors are still anticipated with even smoother bottom/top surfaces and probably with higher material quality.

**C. Results with MBE AlGaAs**

We have so far used the MOCVD epitaxial AlGaAs film in our demonstrations above. Yet another type of epitaxial AlGaAs is grown by MBE. Compared to the MOCVD method, MBE offers advantages in terms of crystal purity and the control of film thickness (down to the monolayer). Therefore, we further explored MBE epitaxial AlGaAs in the waveguide fabrication using in-house MBE growth, in an attempt to attain higher $Q$ factors in microrings. Note that the etch-stop layer is still $Al_{0.8}Ga_{0.2}As$ with a thickness of 500 nm.

With an optimized MBE growth condition in particular for surface quality, we reached the atomic-level smoothness for as-grown AlGaAs surface (bottom of the waveguide) with a roughness of $\sigma = 0.15$ nm, as shown in Fig. 4(a). Figure 4(b) presents the AFM result of the surface after the removal of the etch-stop layer (top of the waveguide), showing a roughness close to the as-grown surface. Both surfaces are further improved compared to MOCVD AlGaAs. We fabricated the microrings with the MBE AlGaAs and then characterized the $Q$ factors, as shown in Fig. 4(c) and (d). For the sake of comparison, we take the results of the microrings with the same design as in Fig. 3(c) and (d). It can be seen that the $Q_0$ is further improved by 43% and 30% for the rings with a radius of 28 μm and 143 μm, respectively. This could be attributed to the improvement of both surface smoothness and material quality. The latter is likely to be a major contribution, regarding that the scattering loss is now dominated by sidewall roughness.

Figure 5 presents the measured $Q_0$ for the AlGaAs rings with different radius and width. The $Q$ is increased by either widening the width or increasing the radius, as the scattering loss is reduced because of the diminished mode interaction with the sidewall roughness. In particular, for the small rings, the $Q$ factor can be effectively improved by increasing the radius. When the radius is above 28 μm and the width is larger than 850 nm, the $Q$ increases very slowly, with the highest value measured to be $3.52 \times 10^6$ in the rings with a radius of 143 μm. To the best of our knowledge, this represents a record $Q$ factor in the integrated III-V microresonators. Apart from the $Q$ factor, another important figure of merit of a cavity is the finesse, which directly measures the field enhancement in the cavity. We obtained a finesse up to $1.4 \times 10^4$ in the ring with a radius of ~12 μm, which is another record result in on-chip III-V microresonators. In such high-finesse cavities, the light-matter interaction can be boosted dramatically, benefiting lots of physical processes, for instance, low-threshold optical parametric oscillations (OPOs) in nonlinear processes.

## D. Summary of $Q$ Measurement and Discussion

Table 1 summarizes the results discussed above, in which we present the measured $Q$ factors for the AlGaAs microrings with different radii, with a width of 850 nm for ~12 μm and ~28 μm radius, and 1000 nm for 143 μm radius. These devices were fabricated from different runs with two epitaxial AlGaAs and different processes, as discussed above. The etch-stop layer is InGaP for Run 1 and Run 2, while $Al_{0.8}Ga_{0.2}As$ for the others. Note that in some comparisons the dimensions of the microrings are slightly different due to different designs used in those runs, and this is also the reason for the lack of some data. Each $Q_0$ factor listed here is the highest measured value over resonances from 1500 nm to 1600 nm in a number of rings with the same dimension. Additionally, in Table 1 we also include the data of Run 4 whose fabrication is the same as Run 5, except for the top surface roughness that is higher than that in Run 5 (due to over etch of the etch-stop layer). By comparing their results, again we show a clear effect of surface roughness on waveguide loss. It is worth noting that, while the surface roughness of the waveguide in Run 4 is >2× higher (i.e., higher surface scattering loss), both Run 3 and Run 4 have very similar $Q$ factors in the rings with the same dimension. This clearly indicates that the MBE AlGaAs performs better than that of MOCVD, regarding to material quality. This is a standard observation resulting from the use of ultrahigh purity elemental source material in MBE as opposed to metalorganic precursors in MOCVD which lead to higher background doping levels.

Table 1 clearly shows that the systematic reduction of surface roughness in AlGaAs-on-insulator waveguides is an effective path for the improvement of $Q$ factors. This can be attained through process optimization including bonding and pattering. In addition to scattering loss, the material loss of epitaxial AlGaAs plays an important role in the final waveguide loss, and MBE AlGaAs offers better material quality in our study. In the measurement, we also observed considerable splitting (on the order of intrinsic linewidth) of resonances, which arises from the scattering-induced inter-mode coupling of counterclockwise and clockwise modes in the microring. As the bottom/top surfaces of the waveguide are nearly atomic-level smooth, the sidewall roughness is believed to be the main scatter of light. This could be further improved by using advanced lithography (e.g., 193 nm DUV) and by further optimizing the etching process. Finally, the surface absorption could be an origin of loss in our AlGaAs waveguide. It has been found that the surface treatment and passivation (with $Al_2O_3$ deposition) can strongly suppress the surface absorption and thus improve $Q$ factors in GaAs resonators [32]. In our waveguide fabrication, $Al_2O_3$ passivation was also implemented on AlGaAs surfaces during processes (before bonding, after etch-stop removal, and after dry etching). For simplicity of fabrication, there was no surface pre-treatment, which may be needed for more effective passivation [32]. Therefore, for further improvement of $Q$ factors, it could be meaningful to comprehensively investigate the impact of surface absorption on AlGaAs waveguide loss in the future.

## 4. FREQUENCY COMB GENERATION

Combining high nonlinearity and high confinement in AlGaAs waveguides, the achieved ultrahigh-$Q$ AlGaAs microrings certainly enable ultrahigh-efficiency nonlinear processes. As a demonstration of such capability, we performed the experiment of frequency comb generation with those AlGaAs rings. The tested rings have a width of ~650 nm with an anomalous group velocity dispersion, and radii of 12 μm (FSR of 1 THz) and 143 μm (FSR of 90 GHz) with the $Q_0$ of ~2×10$^6$ and ~3×10$^6$ at the pump resonances, respectively. Figure 6 shows the output frequency comb spectra for the two rings pumped at different powers. For the 1 THz ring, at the pump power of 23 μW (power in the bus waveguide), the four-wave-mixing sidebands are clearly observed (Fig. 6(a)). At a higher pump power of only 150 μW, the comb lines cover a wavelength range of 150 nm (Fig. 6(b)), while this power is still two times lower than the record OPO threshold in silicon nitride on-chip resonators [20]. For the 90 GHz ring (Fig. 6(c) and (d)), at the pump power of 120 μW, multiple comb lines can be observed. When the pump power increases to 850 μW, the comb extends over a wavelength range of 200 nm; and, what is more, the comb lines are complete over the whole comb span without missing lines. This suggests that the soliton comb could be further reachable at sub-milliwatt power levels. As a repetition rate of 90 GHz is electronically detectable, our result potentially allows for ultralow-power chip-based microcombs in the future. Lastly, we also estimated the threshold power theoretically, using the expression [34] $P_{\text{th}} \approx 1.54(\frac{\pi}{2}) \cdot \frac{Q_c}{2Q_L} \cdot \frac{n^2 V}{n_2 \lambda Q_L^2}$, where $\lambda$ is the pump wavelength, $V$ is the mode volume, and $Q_c$ and $Q_L$ are the coupling and loaded quality factors of the resonator, respectively. Considering a critical coupling and hence $Q_c = 2Q_L = Q_0$, we estimated the threshold powers of ~11 μW and ~54 μW for the 1 THz ring and 90 GHz ring, respectively, which are in rough agreement with the experimental results.

The demonstrated ultralow power threshold in the AlGaAs nanowaveguides can radically lower the operating power for on-chip nonlinear processes. This is of great importance in AlGaAs. First, it dramatically relaxes the requirement for a pump laser needing only a few milliwatts of output power. In contrast, a pump laser with more than tens of milliwatts output power is typically necessary for other nonlinear material platforms, which remains the main limitation for practical applications. Second, because of low operating power, multiple nonlinear functionalities on a single AlGaAs chip become possible; for example, simultaneously integrating second harmonic and frequency comb generation for self-referencing. Finally, AlGaAs PICs are a viable solution to combine with integrated laser technology, such as III-V/Si heterogeneous lasers, towards a monolithic integration of AlGaAs nonlinear PICs.

## 5. CONCLUSION

In conclusion, we significantly reduced losses in AlGaAs-on-insulator waveguides by reducing surface roughness through the systematic optimization of fabrication processes. Based on this, with MBE-grown AlGaAs we demonstrated ultralow-loss high-confinement microring resonators with intrinsic $Q$ factors up to $3.52\times10^6$ and finesse as high as $1.4\times10^4$, which represent record results in the integrated III-V PICs and are even comparable with the state-of-the-art results in SOI PICs. Moreover, thanks to the high nonlinearity and low loss, we showed ultra-efficient frequency comb generations in those AlGaAs resonators and achieved record-low threshold powers on the order of ~20 μW and ~120 μW for the resonators with a 1 THz and a 90 GHz FSR, respectively. These results are important for implementing AlGaAs as a novel integrated nonlinear platform. Our work should stimulate multiple fields of nonlinear research based on second- and third-order nonlinear effects in AlGaAs, such as second-harmonic generation, frequency comb and soliton generation, and all-optical signal processing. It will open new opportunities for chip-based efficiency-demanding nonlinear applications. Finally, combined with existing techniques for (Al)GaAs-related optoelectronics, the currently developed low-loss AlGaAs-on-insulator platform will enable an even wider range of photonic applications with the integration of both passive/active and linear/nonlinear functionalities in (Al)GaAs PICs in the near future.

**Funding**. DARPA MTO DODOS contract (HR0011-15-C-055).

**Acknowledgment**. The authors thank Songtao Liu, Chao Xiang, and Warren Jin for their assistance with fabrication and measurement, and Eric Stanton for useful discussions. The UCSB nano-fabrication facility was used. This research was funded by the Defense Advanced Research Projects Agency (DARPA). The views, opinions and/or findings expressed are those of the author and should not be interpreted as representing the official views or policies of the Department of Defense or the U.S. Government.

**Disclosures.** The authors declare no conflicts of interest.

See Supplement 1 for supporting content.

†These authors contributed equally to this work.

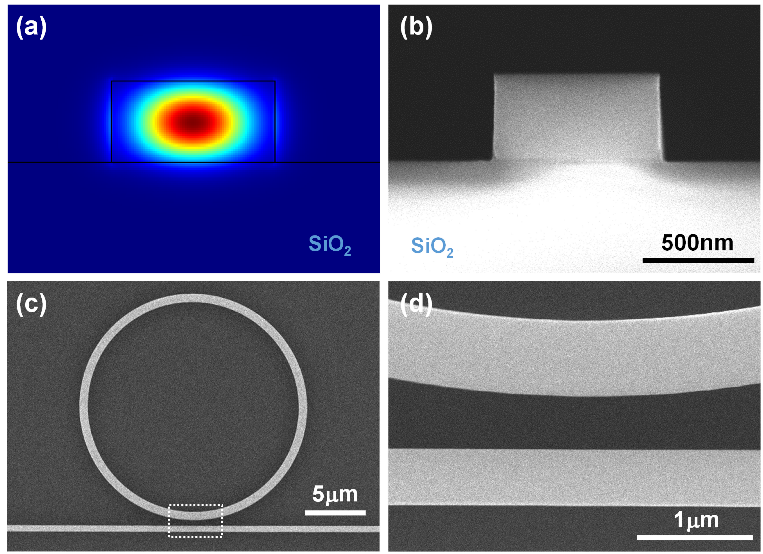

Fig. 1. (a) Simulated fundamental TE mode of AlGaAs waveguide on SiO$_2$ with a height × width of 400 nm × 700 nm. (b) Cross-section SEM image of a fabricated AlGaAs waveguide before oxide cladding. (c) SEM image of an AlGaAs microring coupled with a bus waveguide. (d) Close-up of the coupling region as denoted by the dashed box in (c).

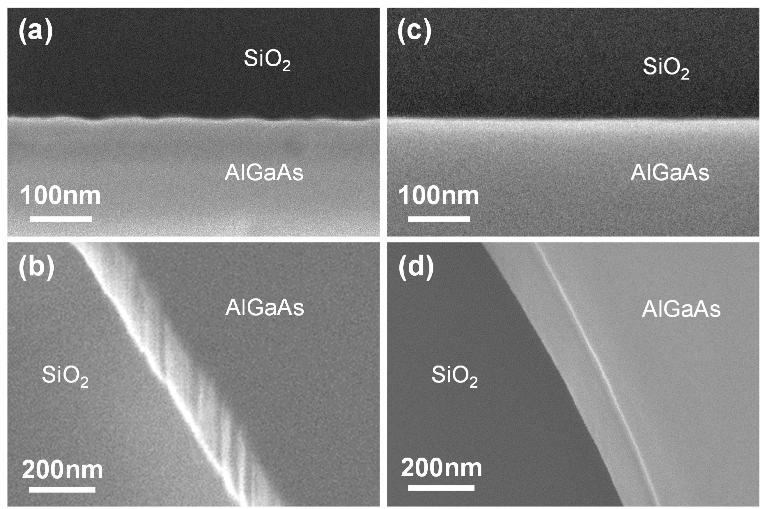

Fig. 2. SEM images of etched AlGaAs waveguide sidewall without/with PR reflow: (a) top-down and (b) tilted views without the reflow; (c) and (d) the corresponding images with the reflow.

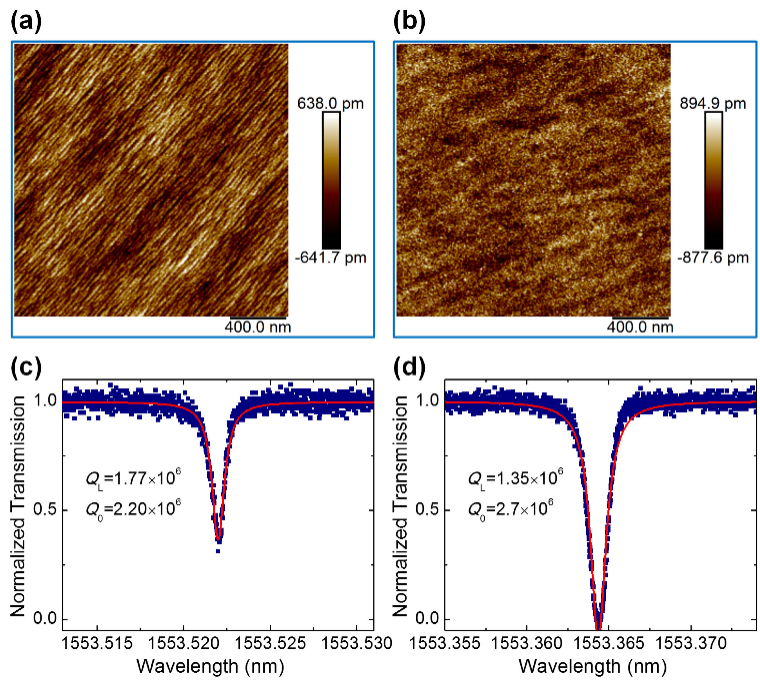

Fig. 3. AFM images of the bottom and top surfaces of waveguide: (a) bottom surface with roughness of $\sigma$ = 0.19 nm; (b) top surface with roughness of $\sigma$ = 0.25 nm. Measured $Q$ factors (loaded $Q_L$ and intrinsic $Q_0$) of the microrings fabricated with MOCVD AlGaAs with improved surfaces: (c) microring with a radius of 28 μm and a width of 850 nm; (d) microring with a radius of 143 μm and a width of 1000 nm. The red lines are fits with a Lorentzian function.

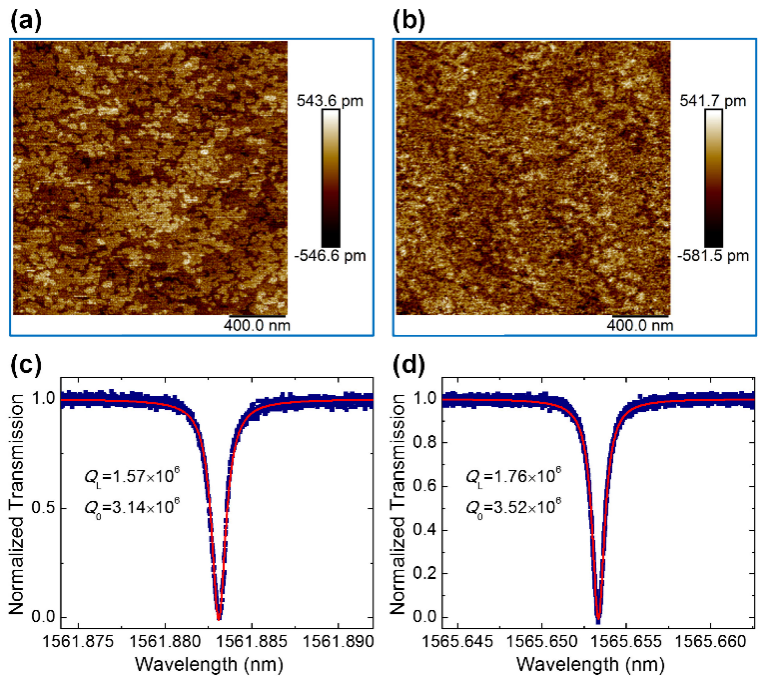

Fig. 4. AFM images of the bottom and top surfaces of waveguide: (a) bottom surface with roughness of $\sigma$ = 0.15 nm; (b) top surface with roughness of $\sigma$ = 0.17 nm. Measured $Q$ factors of the microrings fabricated with MBE AlGaAs with improved surfaces: (c) microring with a radius of 28 µm and a width of 850 nm; (d) microring with a radius of 143 µm and a width of 1000 nm. The red lines are fits with a Lorentzian function.

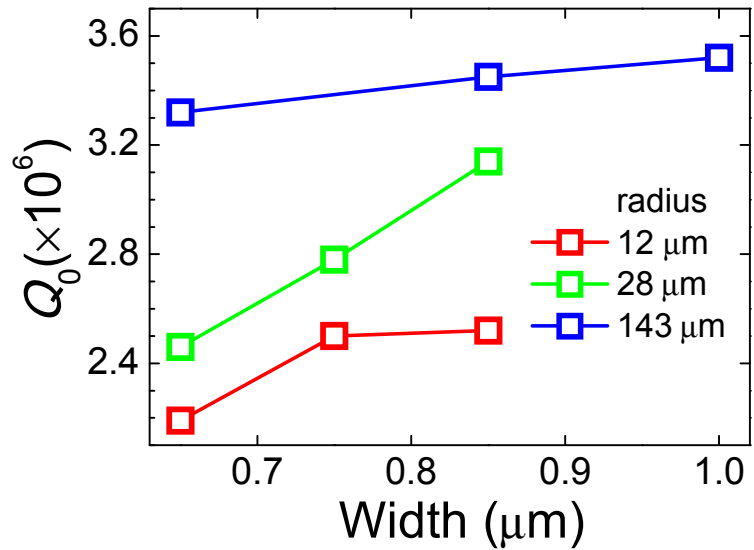

Fig. 5. Measured $Q$ factors for the AlGaAs microrings with different radii and widths.

Table 1. Measured $Q_0$ for the microrings of various radii fabricated from different runs

| Run | AlGaAs | Reflow | Roughness $\sigma$ (nm) | | $Q_0$ (×10$^6$) of rings with various radius | | |
|---|---|---|---|---|---|---|---|
| | | | bottom | top | 12 µm | 28 µm | 143 µm |
| 1 | MOCVD | No | 0.45 | 0.75 | 0.1 | 0.2 | -- |
| 2 | MOCVD | Yes | 0.45 | 0.75 | 0.55 | 1.0 | -- |
| 3 | MOCVD | Yes | 0.19 | 0.25 | 1.53 | 2.2 | 2.7 |
| 4 | MBE | Yes | 0.15 | 0.60 | 1.52 | 2.12 | 2.71 |
| 5 | MBE | Yes | 0.15 | 0.17 | 2.52 | 3.14 | 3.52 |

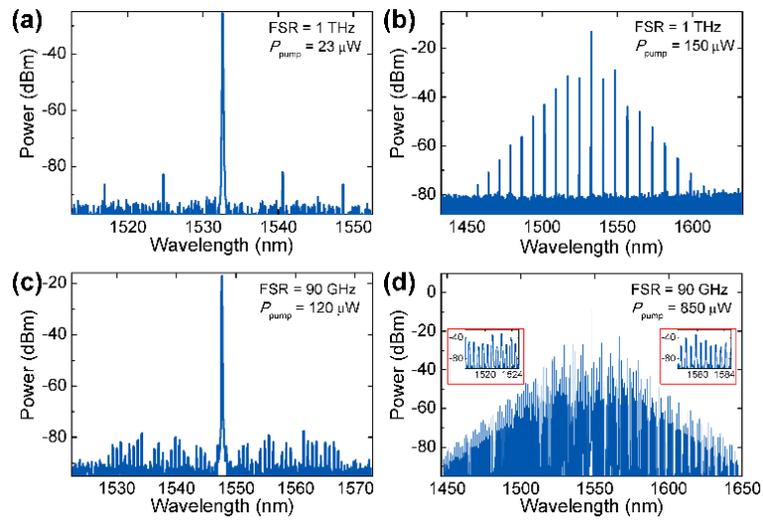

Fig. 6. Frequency comb spectra from two AlGaAs microrings: 1 THz ring at pump powers of (a) 23 µW and (b) 150 µW; 90 GHz ring at pump powers of (c) 120 µW and (d) 850 µW. Insets in (d): close-up of the comb spectra in selected wavelength ranges.

## Supplementary Material

This document provides supplementary information to "Ultrahigh-Q AlGaAs-on-insulator microresonators for integrated nonlinear photonics," including fabrication details and characterization results.

### S1. Fabrication of AlGaAs waveguide

The fabrication of AlGaAs waveguide consists of two steps: wafer bonding and waveguide patterning. We use a heterogeneous wafer-bonding technique as developed previously [1]. Here, we further optimize the fabrication processes, aiming for the loss reduction of the waveguide. Figure S1 shows the fabrication flow.

The fabrication begins with the substrate preparation: AlGaAs epitaxial wafer and silicon substrate with a 3 μm thermal $SiO_2$ box layer on top. For the epitaxial wafer, a 500nm-thick etch-stop layer was grown first on the (100) surface of a GaAs wafer, followed by a 400nm-thick AlGaAs. In our fabrication, two types of the etch-stop layer were used – InGaP and $Al_{0.8}Ga_{0.2}As$, and both MOCVD and MBE growth methods were employed. Note that after growth, a 6nm-thick $Al_2O_3$ layer was finally deposited on the AlGaAs surface by atomic layer deposition (ALD) for possible passivation. The $SiO_2$ box on the silicon substrate was patterned by etching 3μm-deep cross-shape trenches using inductively coupled plasma reactive-ion etching (ICP-RIE) with the chemistry of $CHF_3$. Those trenches are used for effective outgassing during bonding. After patterning, the silicon wafer was cleaned for bonding. Before bonding, the surfaces of two substrates were activated in $O_2$ plasma. The bonding was carried out at 100 °C for 24 hours with a ~1 MPa pressure applied between two substrates to enhance the bonding strength.

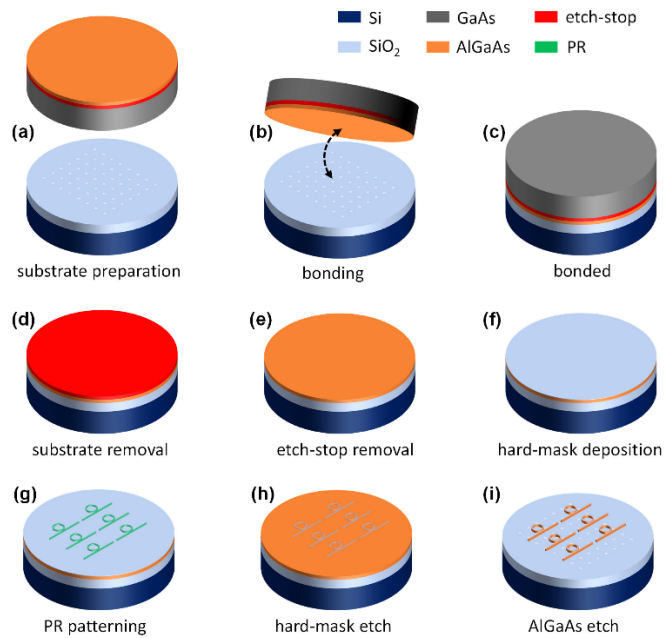

Fig. S1. (a-i) Sketch of fabrication flow for AlGaAs-on-insulator waveguide.

After bonding, the GaAs substrate was removed in a mixture of $H_2O_2$ and $NH_4OH$ at room temperature. First, a volume ratio of $H_2O_2 : NH_4OH = 10 : 1$ was used to etch most of GaAs, with a ~100 μm GaAs left. Then, a volume ratio of $H_2O_2:NH_4OH=30:1$ was used to remove the rest GaAs. The etch-stop layer was selectively etched in HCl (37%) in the case of InGaP, or in a diluted HF (HF(50%) : DI = 1 : 20, in volume) in the case of $Al_{0.8}Ga_{0.2}As$. The final surface of AlGaAs was characterized by using atomic force microscope (AFM). Figure S2 shows the picture of the bonded AlGaAs film on the silicon substrate.

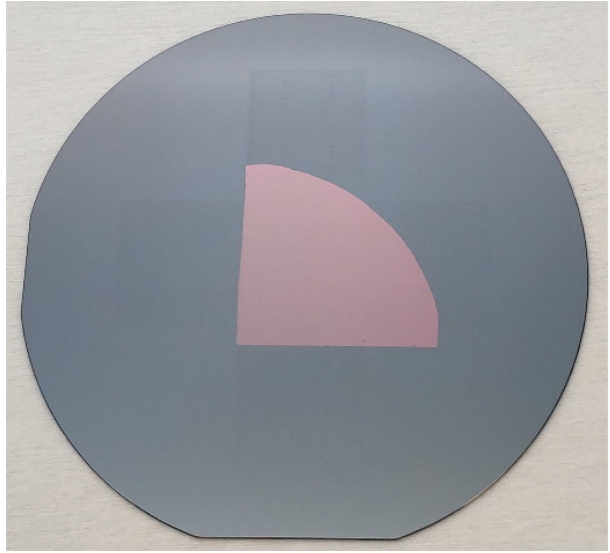

Fig. S2. Bonded a quarter of 3-inch AlGaAs on a 4-inch Si wafer.

For waveguide patterning, we deposited 6 nm $Al_2O_3$ for passivation and 100 nm $SiO_2$ for the hard mask by ALD at 200 °C, respectively. The wafer was then patterned using the photoresist (PR) of UV6™-0.8 with a thickness of ~600 nm. Prior to the PR, an anti-reflective (AR) coating (DUV-42P) (~60 nm) was applied. After the development of PR, a thermal reflow was carried out at 155 °C for 3 min using a hotplate. Afterward, the AR was etched by ICP-RIE with $O_2$, and then the $SiO_2$ hard mask was partially etched with gases of $CHF_3/CF_4/O_2$ in the same ICP-RIE tool. Here, a two-step-etch process was developed for the hard mask. After the first etch, the wafer was cleaned to remove any organic residues and avoid unwanted pattern transfers in the final waveguides. The rest of the hard mask was directly etched with the same dry etch process. The AlGaAs was etched by ICP-RIE with gases of $Cl_2/N_2$. Figure S3 shows the scanning electron microscope (SEM) images of the etched AlGaAs microrings. Finally, the sample was deposited with a ~8 nm $Al_2O_3$ by ALD and then cladded with a ~1.5 μm $SiO_2$ by using plasma-enhanced chemical vapor deposition (PECVD).

In the fabrication, we employed a 248nm deep-ultraviolet (DUV) stepper for lithography on a 100 mm wafer. The UCSB Nanofabrication Facility was used for all processes [2].

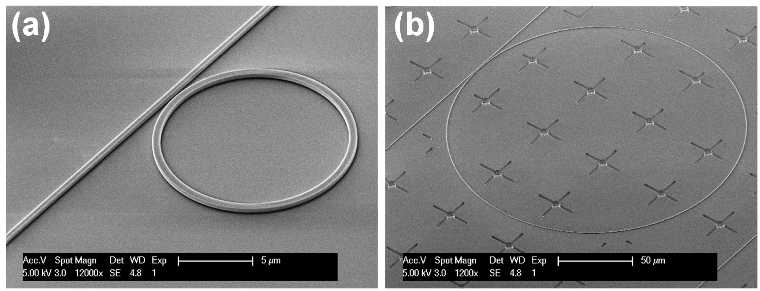

Fig. S3. SEM of images of two microrings with different radius: (a) 13 µm; (b) 143 µm.

**S2. The reflow of PR patterns**

The reflow of PR is an effective way to reduce the line-edge roughness of PR patterns, ultimately avoiding roughness transfer into waveguides during dry etching. Figure S4 shows the SEM images of PR patterns without/with the reflow process. It can be clearly seen that the reflow can significantly smoothen the edges of PR patterns. Therefore, the roughness transferred to waveguide sidewalls during dry etching will be minimized, eventually resulting in smooth sidewalls.

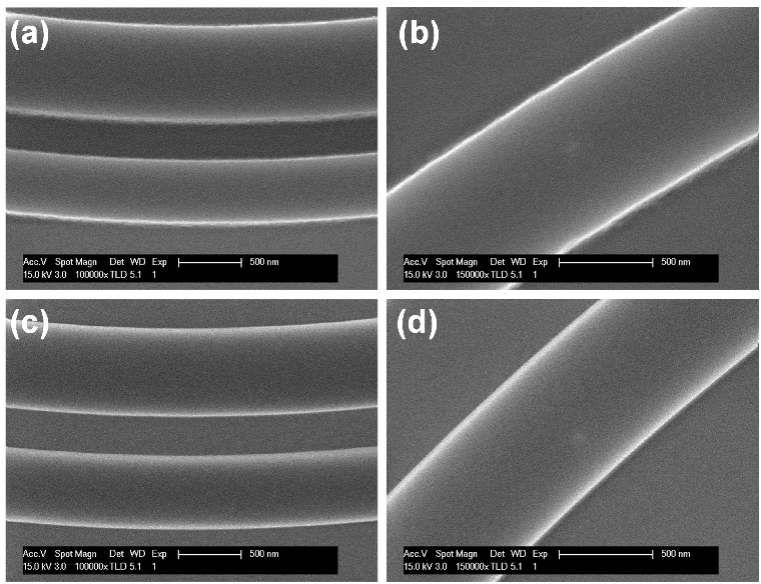

Fig. S4. SEM results of PR patterns: (a) and (b) without reflow; (c) and (d) with reflow.

**S3. Roughness with InGaP etch-stop layer**

Figure S5 shows the AFM results of MOCVD-grown AlGaAs with an etch-stop layer of InGaP. Both the as-grown surface and the surface after the removal of InGaP have a higher root-mean-square (RMS) roughness, compared to that with $Al_{0.8}Ga_{0.2}As$ as the etch-stop layer.

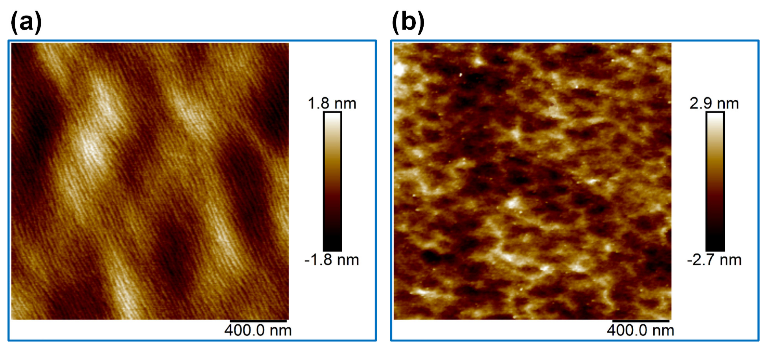

Fig. S5. AFM images of AlGaAs with an etch-stop layer of InGaP. (a) The as-grown surface of AlGaAs with a RMS roughness of ~0.45 nm. (b) The AlGaAs surface after the removal of InGaP with a RMS roughness of ~0.75 nm.